# Epitaxial growth and properties of $La_{0.7}Sr_{0.3}MnO_3$ thin films with micrometer wide atomic terraces

Wei Yuan[1], Yuelei Zhao[1], Chi Tang[2], Tang Su[1], Qi Song[1], Jing Shi[2,a)], and Wei Han[1,3,b)]

[1]International Center for Quantum Materials, Peking University, Beijing, 100871, P. R. China

[2]Department of Physics and Astronomy, University of California, Riverside, California 92521, USA

[3]Collaborative Innovation Center of Quantum Matter, Beijing 100871, P. R. China

**Abstract:**

$La_{0.7}Sr_{0.3}MnO_3$ (LSMO) films with extraordinarily wide atomic terraces are epitaxially grown on $SrTiO_3$ (100) substrates by pulsed laser deposition. Atomic force microscopy measurements on the LSMO films show that the atomic step is ~ 4 Å and the atomic terrace width is more than 2 μm. For a 20 monolayers (MLs) LSMO film, the magnetization is determined to be 255 ± 15 emu/cm$^3$ at room temperature, corresponding to 1.70 ± 0.11 $\mu_B$ per Mn atom. As the thickness of LSMO increases from 8 MLs to 20 MLs, the critical thickness for the temperature dependent insulator-to-metal behavior transition is shown to be 9 MLs. Furthermore, post-annealing in oxygen environment improves the electron transport and magnetic properties of the LSMO films.

a) Email: jing.shi@ucr.edu

b) Email: weihan@pku.edu.cn

La$_{0.7}$Sr$_{0.3}$MnO$_3$ (LSMO) is a very attractive material for spintronics due to the experimental observation of the half-metallic ferromagnetism [1] and the relatively high Curie temperature ($T_C$~ 369 K) compared with common oxide-based magnetic materials [2]. For example, LSMO has been widely used for efficient spin injection in organic spin valves [3-8] and giant tunneling magnetoresitance in magnetic tunnel junctions [9,10]. Regarding the synthesis of thin LSMO films, pulsed laser deposition (PLD) has been found to be a promising method to achieve high quality films with atomic scale variations [11-14]. Common to strongly correlated oxide materials, the intriguing magnetic and electron transport properties of LSMO are intimately related to the spin-lattice-charge couplings, which are sensitive to strain, growth condition, and the thickness of the films [14,15]. For example, the magnetic anisotropy of LSMO has been shown to be in-plane when it is grown on SrTiO$_3$ (STO) substrates but out-of plane on LaAlO$_3$ (LAO) substrates [16,17]. Strain effect has also been found to play a critical role in the electron transport properties via growing LSMO on various perovskites, including STO, LAO, DyScO$_3$, and NdGaO$_3$ [18,19]. Furthermore, the magnetic and electron transport properties of the LSMO films are also shown to be highly dependent on the growth condition such as the oxygen pressure [11,12]. As single terraces of ferromagnetic films with desired spin structures have the potential to create future spintronics devices, the atomic level control of the magnetic oxide thin films with wide atomic terraces is emergent. However, in previous reports, the terraces of these LSMO films are only several hundred nanometers wide [12,15] and the study of epitaxial LSMO films with micrometer wide atomic terraces has been lacking.

In this letter, we report the growth of LSMO films with extraordinarily wide atomic terraces of more than 2 μm on STO (100) via PLD and the characterization of the magnetic and electron transport properties. *In-situ* reflected high energy electron diffraction (RHEED) and *ex-situ* x-ray diffraction (XRD) indicate the epitaxial growth of the crystalline LSMO films. The saturated magnetization for a 20 monolayers (MLs) film is determined to be 255 ± 15 emu/cm$^3$ at room temperature, corresponding to 1.70 ± 0.11 $\mu_B$ per Mn atom. The temperature dependence of the sheet resistance for 8 - 20 MLs LSMO films indicates a critical thickness of 9 MLs for the insulator-to-metal transition. We further demonstrate the ability to improve the electron transport and magnetic properties of LSMO films via post-annealing in oxygen environment. The ability to control the ultrathin film growth with atomic level precision is very important for the research fields of current interest, such as oxide spintronics and organic spintronics, and could lead to future spintronics devices made from single terraces of ferromagnetic or antiferromagnetic films with desired spin structures.

The LSMO films are grown using a KrF laser with the wavelength of 248 nm and the energy between 200 and 220 mJ. The repetition rate of the laser is 5 Hz with a pulse duration of 30 ns. Prior to the LSMO growth, the commercial STO (100) substrates, are annealed *in-situ* at 700 °C in 0.03 mbar oxygen environment for ~30 min to form ultra-flat surface with atomic terraces. Figures 1a and 1b show the RHEED pattern and atomic force microscopy (AFM) image of a typical STO substrate after this annealing process. Then, the LSMO target is cleaned by the laser with 1000 - 2000 pulses right before opening the shutter. During the growth, the substrate temperature is kept at 700 °C; the partial pressure of the oxygen is 0.03 mbar; and the distance between the LSMO target and the STO substrates is 55 mm. The RHEED intensity in the red square (shown in figure 1a) is monitored and the oscillations of RHEED intensity demonstrate the layer by layer growth mode of the LSMO. One representative RHEED intensity oscillation curve for the growth of a 10 MLs LSMO film is shown in figure 1c. At ~35 seconds, the shutter is open and the LSMO is deposited on top of the STO substrate. At ~163 seconds, the RHEED intensity shows a minimum value, which indicates the growth of an half ML LSMO film. Then the RHEED intensity starts to increase and reaches a maximum value at ~249 seconds and this indicates the growth of a full ML film. At ~1820 seconds, the shutter is closed and the measurement of the RHEED intensity is stopped. The total 10 peaks of the RHEED intensity oscillation curve indicate the growth of exact 10 MLs LSMO film on the top of the STO substrate. After the growth, the RHEED pattern of the 10 MLs LSMO film at 700 °C is shown in figure 1d. The RHEED pattern is very sharp and streaky, indicating the crystalline structure of the epitaxially grown LSMO film. Then, the substrate temperature is decreased down to 300 °C with a rate of 5 °C/min in the PLD chamber with the oxygen partial pressure of 0.03 mbar, which is the same as that during the growth. When the temperature reaches 300 °C, the substrate heater is turned off to cool down the samples naturally. As the temperature is lower than 70 °C, the oxygen is pumped out from the PLD chamber and the LSMO samples are transferred out of the chamber for further *ex-situ* characterization.

The surface morphology of epitaxially grown LSMO films is characterized by a Bruker AFM system. The films are found to exhibit very wide terraces from ~2 to ~10 μm. Figure 1e shows an AFM image measured on a typical 10 MLs LSMO film. One line cut across the terraces in this AFM image is plotted in figure 1f. The step height is ~4 Å, which is equal to the height of one ML LSMO film in (100) direction. The terraces are ~3 μm wide, which are extraordinarily wide compared to several hundred nm in previous reports [12,15]. We believe that such extraordinarily wide terraces in LSMO can be attributed to the wide terraces of the annealed STO substrates prior to the LSMO growth (figure 1b). To further confirm it, two STO substrates are loaded into the PLD chamber and they undergo the identical annealing process. Then one STO substrate is taken out and the surface morphology is measured by AFM. For the second

STO substrate, we grow 10 MLs LSMO film on top of the STO substrate and then the film is measured by AFM. Based the AFM results measured on multiple spots across the sample surface, the terrace width of both the STO substrate and the LSMO film ranges from ~3 to ~10 µm, supporting that the LSMO films are epitaxially grown on terraced STO substrates and the wide terraces of the LSMO films originate from the wide terraces formed on the surface of STO substrates after annealing in the PLD chamber.

The crystalline structure and the magnetic properties of these LSMO films with extraordinarily wide terraces are characterized using a high resolution Bruker Discovery D8 system and a Lakeshore 7400 vibrating sample magnetometer respectively. Figure 2a shows the XRD data of the θ-2θ scans on the 10, 16, and 20 MLs films. The peak at ~46.5 degrees corresponds to the STO (200). For 10 MLs LSMO films (black curves), the intensity of the LSMO signal is rather weak. As the thickness increases, the intensity of the LSMO (200) feature also increases. These results further prove the crystalline nature of LSMO films. The magnetic properties of the LSMO films are characterized by a vibrating sample magnetometer at room temperature. Figure 2b shows the magnetization *vs.* in plane magnetic field for a 20 MLs LSMO sample. The room temperature saturated magnetization ($M_S$) is calculated to be 255 ± 15 emu/cm$^3$, corresponding to 1.70 ± 0.11 $\mu_B$ per Mn atom. This $M_S$ value is similar to that of a 28 nm LSMO sample (~70 MLs) obtained at room temperature reported previously [11].

The electron transport properties of these epitaxial grown LSMO films are measured in a closed-cycle refrigerator system with the temperature ranging from 300 to ~16 K and/or a Quantum Design physical property measurement system (PPMS) with the temperature ranging from 350 to 10 K. The sheet resistance is characterized by the standard van der Pauw method with four copper wires connected to the corners of the LSMO films with Ag paint for electrical connection, which is shown in the inset of figure 3b. The resistivity (ρ) is calculated based on the resistances obtained for four deferent geometries using the following formula:

$$\rho = \frac{d\pi}{\ln 2}\left(\frac{R_{12,43} + R_{23,14} + R_{34,21} + R_{41,32}}{4}\right) \quad (1)$$

in which $d$ is the thickness of the LSMO films, $R_{12,43}$, $R_{23,14}$, $R_{34,21}$ and $R_{41,32}$, are the resistance values obtained using the geometry by passing current between contacts *1* and *2*, *2* and *3*, *3* and *4*, *4* and *1*, and measuring the voltage between contacts *4* and *3*, *1* and *4*, *2* and *1*, *3* and *2*, respectively." $R_{12,43}$ is measured using the geometry in the inset of figure 3b. The DC current (*I*) across the films is +1 µA and -1 µA, provided by a Keithley 6221 AC/DC current source, and the voltage is determined by averaging between the two voltage values measured when $I = +1$ µA and $I = -1$ µA via a Keithley 2182 nanovoltmeter. We study the electron transport properties for different thicknesses of the LSMO films (8 - 20 MLs), which are all determined by the RHEED intensity oscillations. Figure 3a shows three typical curves of the RHEED intensity oscillations for 8, 11, and 20 MLs LSMO films. The resistivity in zero magnetic field *vs.* temperature is shown in figure 3b. For the 8 MLs LSMO film, the resistivity increases as the temperature decreases below 150 K, which is a semiconducting behavior. For the 9 to 20 MLs films, the resistivity decreases as the temperature decreases, which is a metallic behavior. For 20 MLs films, the $T_P$ is ~ 342 K, which is close to Curie temperature for bulk LSMO (~369 K), indicating the high quality of the LMSO films. These results show that the critical thickness for insulator-to-metal transition is 9 MLs, which is significantly smaller than ~80 Å (~ 20 MLs) reported previously [20] and indicates the high quality of the epitaxially grown LSMO films in our study. It is also noticed that there is a resistance peak at ~ 200 K for 8 MLs LSMO film, which has been shown to be related to the ferromagnetic to paramagnetic transition [14,20,21].

As the thickness increases, this resistivity peak temperature ($T_p$) increases, which indicates a higher $T_C$ for thicker LSMO films.

Finally, we further improve the properties of the LSMO films by post-annealing in oxygen environment with oxygen partial pressure from 3 to 500 mbar. Upon the film growth in 0.03 mbar as discussed above, we increase the oxygen pressure up to 3 - 500 mbar before cooling down the samples. We expect this post-annealing with much higher oxygen pressure to remove oxygen vacancies which will result in even better film quality. The electron transport properties of the 10 MLs films annealed in 0.03, 3, 50, 200 and 500 mbar are shown in figure 4a, respectively. Indeed, the post-annealing in higher pressure oxygen environment makes the LSMO films more metallic and the $T_p$ higher. The $T_p$ vs. oxygen pressure is summarized in figure 4b. As the oxygen pressure increases from 0.03 to 50 mbar, the $T_p$ increases from ~270 K up to ~310 K. Then the $T_p$ saturates at ~310 K with the oxygen pressure increasing further up to 500 mbar, which can be attributed to the fact that very few oxygen vacancies exist in the LSMO films after post-annealing [14].

In summary, ultrathin LSMO films with extraordinarily wide terraces are grown on STO (100) using PLD. The critical thickness for the temperature dependent insulator-to-metal behavior transition of grown LSMO films is shown to be 9 MLs. Furthermore, it is demonstrated that the electron transport and magnetic properties of the LSMO films could be further improved via post-annealing and room temperature metallicity is observed for 10 MLs films. These results are very important for oxide spintronics and organic spintronics, which are two research fields of current interest. For example, one can fabricate the organic spin valve, or magnetic tunnel junction on a single terrace of the LSMO films without any atomic steps. Previous works on LSMO based organic spin valve and magnetic tunnel junctions all include multiple atomic steps, which could play an important role for spin current injection. If we remove the atomic steps and use only one single terrace, the effect of these atomic steps on the magnetoresistance will be resolved. Besides, our results could lead to future spintronics devices made from single terraces of ferromagnetic or antiferromagnetic films with desired spin structures. These wide terraces provide an opportunity of integrating an atomic 2D material, such as graphene with the single terrace of ferromagnetic and/or antiferromagnetic materials without any atomic steps. Recently, the proximity effect between graphene and magnetic films has attracted a lot interest for the search of quantum anomalous Hall effect in graphene based structures. It has been reported that graphene could be magnetic when it is put on YIG, a ferromagnetic insulator [22]. However, there were terraces of several hundred nm for the YIG. As a result of which, graphene cannot be put flatly on top of the YIG that might limit the properties of the magnetic graphene. Our achievement of very wide terraces could provide an ideal platform to study the proximity effect and quantum anomalous Hall effect between graphene and the LSMO films. Furthermore, it is known that the atomic level control of material growth can lead to physical phenomena and a better understanding of the phenomena in the material. Another intriguing point is that our work demonstrated the possibility of very wide terraces, which will motivate a lot of materials scientists to study and improve various ferromagnetic and antiferromagnetic materials so that these materials could have wide terraces, which could lead to interesting properties.

## ACKNOWLEDGEMENT

We acknowledge the support of National Basic Research Programs of China (973 Program). Wei Han also acknowledge the support by the 1000 Talents Program for Young Scientists of China.

**REFERENCES:**

[1] J. H. Park, E. Vescovo, H.-J. Kim, C. Kwon, R. Ramesh, and T. Venkatesan, Direct evidence for a half-metallic ferromagnet, Nature **392**, 794 (1998).

[2] A. Urushibara, Y. Moritomo, T. Arima, A. Asamitsu, G. Kido, and Y. Tokura, Insulator-metal transition and giant magnetoresistance in $La_{1-x}Sr_xMnO_3$, Phys. Rev. B **51**, 14103 (1995).

[3] Z. H. Xiong, D. Wu, Z. Valy Vardeny, and J. Shi, Giant magnetoresistance in organic spin-valves, Nature **427**, 821 (2004).

[4] L. E. Hueso, J. M. Pruneda, V. Ferrari, G. Burnell, J. P. Valdes-Herrera, B. D. Simons, P. B. Littlewood, E. Artacho, A. Fert, and N. D. Mathur, Transformation of spin information into large electrical signals using carbon nanotubes, Nature **445**, 410 (2007).

[5] D. Sun, L. Yin, C. Sun, H. Guo, Z. Gai, X. G. Zhang, T. Ward, Z. Cheng, and J. Shen, Giant Magnetoresistance in Organic Spin Valves, Phys. Rev. Lett. **104**, 236602 (2010).

[6] C. Barraud, P. Seneor, R. Mattana, S. Fusil, K. Bouzehouane, C. Deranlot, P. Graziosi, L. Hueso, I. Bergenti, V. Dediu, F. Petroff, and A. Fert, Unravelling the role of the interface for spin injection into organic semiconductors, Nat. Phys. **6**, 615 (2010).

[7] F. Li, T. Li, and X. Guo, Vertical Graphene Spin Valves Based on $La_{2/3}Sr_{1/3}MnO_3$ Electrodes, ACS Applied Materials & Interfaces **6**, 1187 (2014).

[8] F. Wang and Z. V. Vardeny, Recent advances in organic spin-valve devices, Synthetic Metals **160**, 210 (2010).

[9] Y. Lu, X. W. Li, G. Q. Gong, G. Xiao, A. Gupta, P. Lecoeur, J. Z. Sun, Y. Y. Wang, and V. P. Dravid, Large magnetotunneling effect at low magnetic fields in micrometer-scale epitaxial $La_{0.67}Sr_{0.33}MnO_3$ tunnel junctions, Phys. Rev. B **54**, R8357 (1996).

[10] M. Bowen, M. Bibes, A. Barthélémy, J.-P. Contour, A. Anane, Y. Lemaître, and A. Fert, Nearly total spin polarization in $La_{2/3}Sr_{1/3}MnO_3$ from tunneling experiments, Appl. Phys. Lett. **82**, 233 (2003).

[11] M. Huijben, L. Martin, Y. H. Chu, M. Holcomb, P. Yu, G. Rijnders, D. Blank, and R. Ramesh, Critical thickness and orbital ordering in ultrathin La0.7Sr0.3MnO3 films, Phys. Rev. B **78**, 094413 (2008).

[12] J. H. Song, T. Susaki, and H. Y. Hwang, Enhanced Thermodynamic Stability of Epitaxial Oxide Thin Films, Advanced Materials **20**, 2528 (2008).

[13] J. X. Ma, X. F. Liu, T. Lin, G. Y. Gao, J. P. Zhang, W. B. Wu, X. G. Li, and J. Shi, Interface ferromagnetism in (110)-oriented $La_{0.7}Sr_{0.3}MnO_3/SrTiO_3$ ultrathin superlattices, Phys. Rev. B **79**, 174424 (2009).

[14] M. Cesaria, A. P. Caricato, G. Maruccio, and M. Martino, LSMO – growing opportunities by PLD and applications in spintronics, J. Phys.: Conf. Ser. **292**, 012003 (2011).

[15] H. Boschker, M. Huijben, A. Vailionis, J. Verbeeck, S. v. Aert, M. Luysberg, S. Bals, G. v. Tendeloo, E. P. Houwman, G. Koster, D. H. A. Blank, and G. Rijnders, Optimized fabrication of high-quality $La_{0.67}Sr_{0.33}MnO_3$ thin films considering all essential characteristics, J. Phys. D: Appl. Phys. **44**, 205001 (2011).

[16] C. Kwon, M. C. Robson, K. C. Kim, J. Y. Gu, S. E. Lofland, S. M. Bhagat, Z. Trajanovic, M. Rajeswari, T. Venkatesan, A. R. Kratz, R. D. Gomez, and R. Ramesh, Stress-induced effects in epitaxial $(La_{0.7}Sr_{0.3})MnO_3$ films, J. Magn. Magn. Mater. **172**, 229 (1997).

[17] J. Z. Sun, D. W. Abraham, R. A. Rao, and C. B. Eom, Thickness-dependent magnetotransport in ultrathin manganite films, Appl. Phys. Lett. **74**, 3017 (1999).

[18]  Y. Takamura, R. V. Chopdekar, E. Arenholz, and Y. Suzuki, Control of the magnetic and magnetotransport properties of $La_{0.67}Sr_{0.33}MnO_3$ thin films through epitaxial strain, Appl. Phys. Lett. **92**, 162504 (2008).

[19]  B. Wang, L. You, P. Ren, X. Yin, Y. Peng, B. Xia, L. Wang, X. Yu, S. Mui Poh, P. Yang, G. Yuan, L. Chen, A. Rusydi, and J. Wang, Oxygen-driven anisotropic transport in ultra-thin manganite films, Nat. Commun. **4**, 2778 (2013).

[20]  M. Angeloni, G. Balestrino, N. G. Boggio, P. G. Medaglia, P. Orgiani, and A. Tebano, Suppression of the metal-insulator transition temperature in thin $La_{0.7}Sr_{0.3}MnO_3$ films, J. Appl. Phys. **96**, 6387 (2004).

[21]  A. P. Ramirez, Colossal magnetoresistance, J. Phys.: Condens. Matter **9**, 8171 (1997).

[22]  Z. Wang, C. Tang, R. Sachs, Y. Barlas, and J. Shi, Proximity-Induced Ferromagnetism in Graphene Revealed by the Anomalous Hall Effect, Phys. Rev. Lett. **114**, 016603 (2015).

**FIGURE CAPTIONS:**

Figure 1: *In-situ* RHEED characterization and *ex-situ* AFM measurement of the STO (100) substrates and the LSMO thin films. (a) The RHEED pattern of STO at 700 °C prior to the LSMO growth. (b) AFM images of a typical STO substrate after high temperature annealing in the PLD chamber (c) A representative RHEED intensity oscillation curve for the growth of a 10 MLs LSMO film. (d) The RHEED pattern of a 10 MLs LSMO thin film at 700 °C. (e) AFM images of a typical 10 MLs LSMO film. (f) An AFM line cut across the terraces in the figure (e).

Figure 2: The structure and magnetic properties of the LSMO films epitaxially grown on STO substrates. (a) θ-2θ scans of the 10, 16 and 20 MLs LSMO films measured by XRD. (b) Magnetization measurement on a 20 MLs LSMO film performed at room temperature with the in-plane magnetic field from -5000 Oe to +5000 Oe using a vibrating sample magnetometer.

Figure 3: Electron transport properties of the LSMO films from 8 to 20 MLs. (a) The RHEED intensity oscillation curves for 8, 11 and 20 MLs LSMO films respectively. (b) The resistivity *vs.* temperature for LSMO films from 8 to 20 MLs. The inset shows the van der Pauw measurement geometry with current following from contact *1* to contact *2* and voltage measured between contacts *4* and *3*.

Figure 4: Improvement of the electron transport properties by post-annealing. (a) The resistivity *vs.* temperature for LSMO films after post-annealing with oxygen pressure from 0.03 to 500 mbar. (b) $T_P$ as a function of the oxygen partial pressure during the post-annealing process.

Figure 1

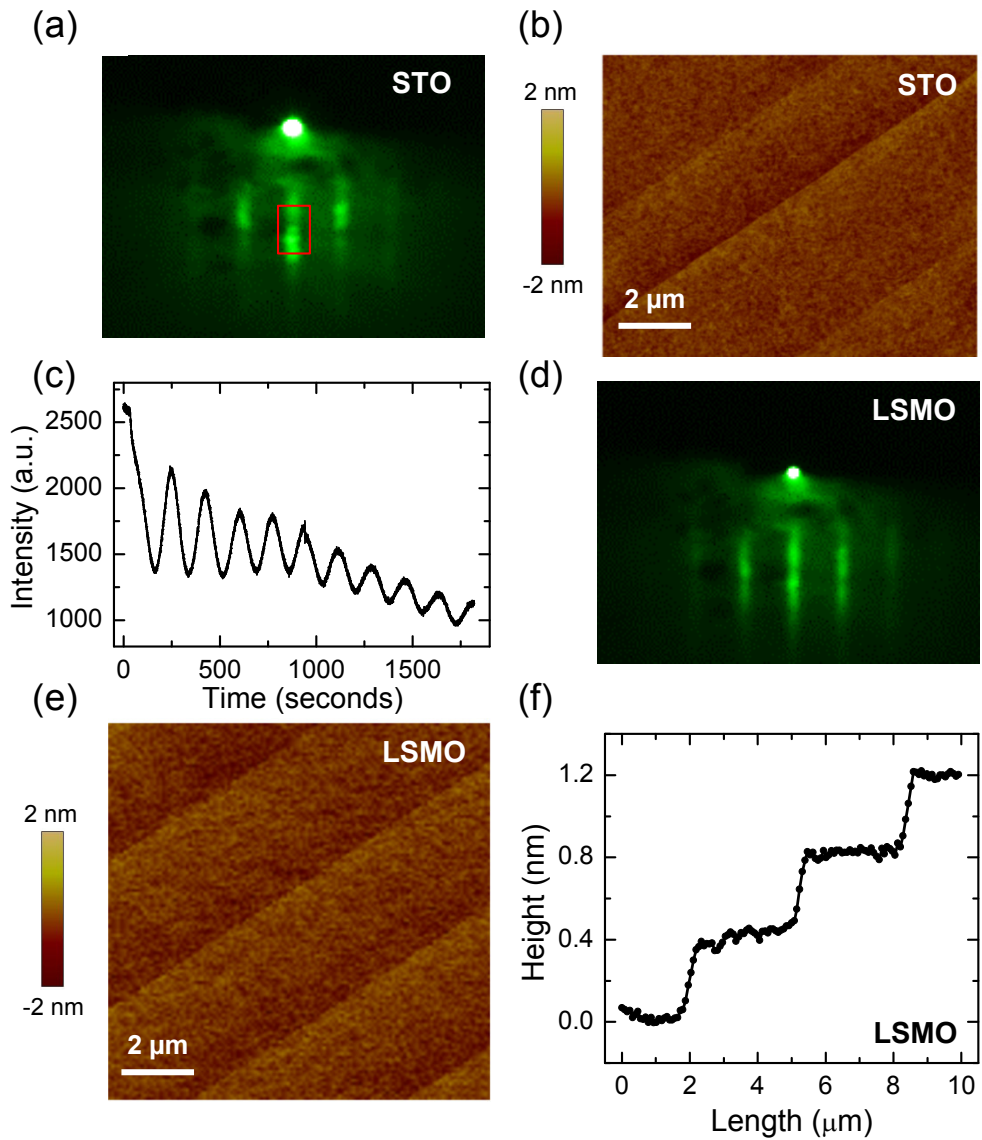

Figure 2

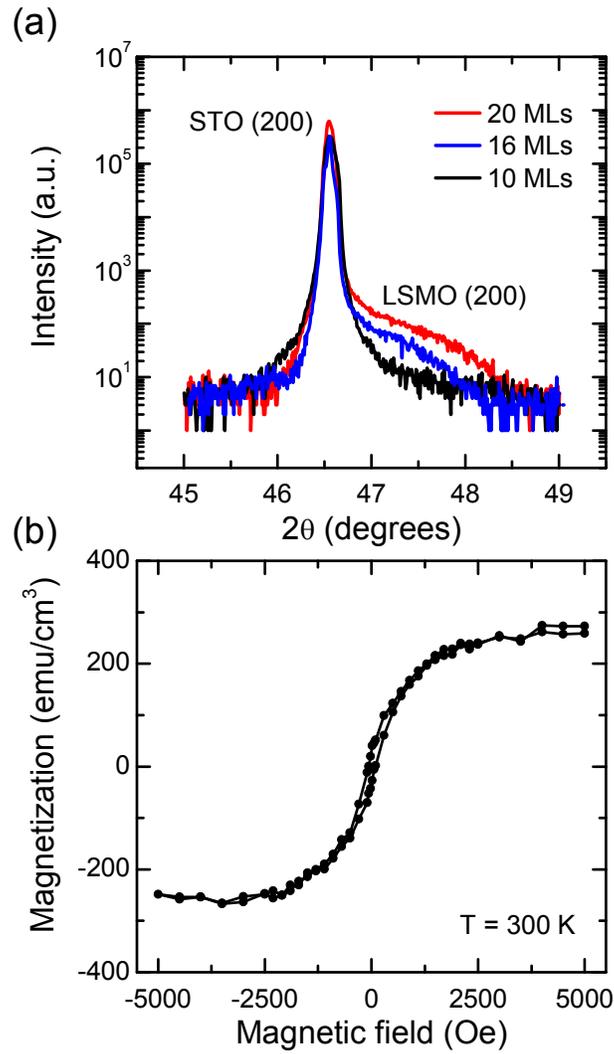

Figure 3

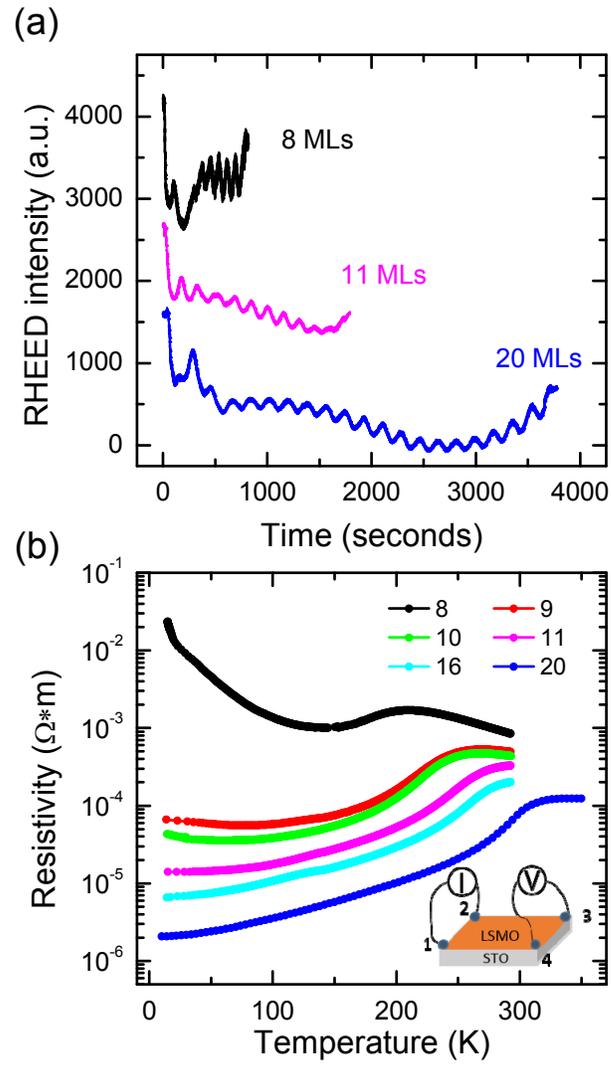

Figure 4

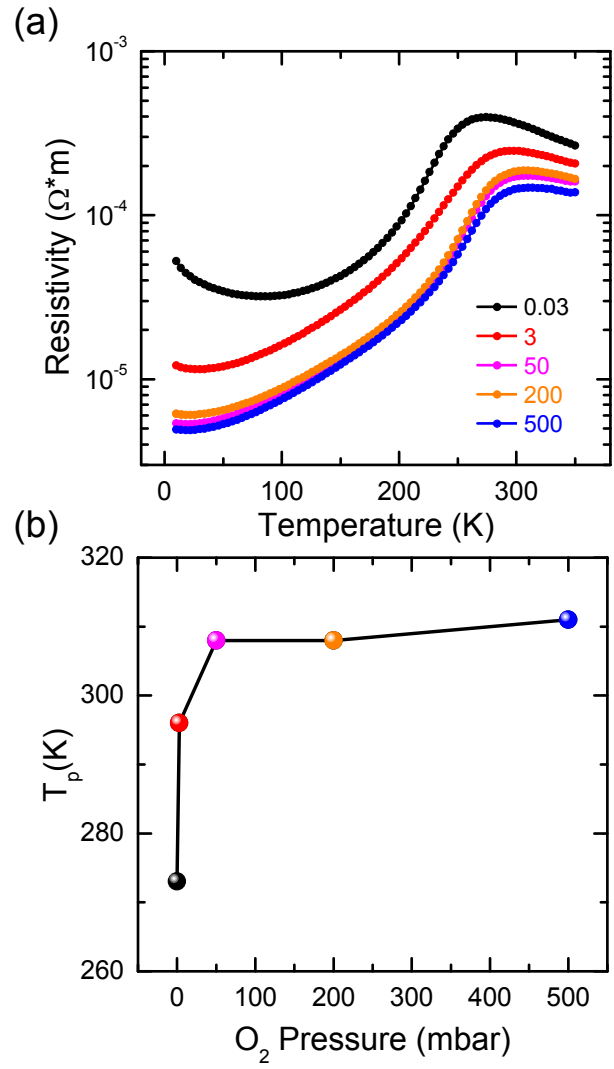